# *Non-localized states and high hole mobility in amorphous germanium*

Tuan T. Tran [1*], Jennifer Wong-Leung [1], Lachlan A. Smillie [1], Anders Hallén [2], Maria G. Grimaldi [3] and Jim S. Williams [1]

*[1] Department of Electronic Materials Engineering, Research School of Physics and Engineering, Australian National University, Canberra, ACT 0200, Australia.*

*[2] KTH — Royal Institute of Technology, School of Information and Communication Technology, PO Box 229, SE-164 40 Kista, Sweden.*

*[3] CNR-IMM MATIS and Department of Physics and Astronomy, University of Catania, Via S. Sofia 64, I-95123 Catania, Italy.*

*(*) corresponding author: tranthientuan@gmail.com*

Covalent amorphous semiconductors, such as amorphous silicon (a-Si) and germanium (a-Ge), are commonly believed to have localized electronic states at the top of the valence band and the bottom of the conduction band. Electrical conductivity is thought to be by the hopping mechanism through localized states. The carrier mobility of these materials is usually very low, in the order of $\sim 10^{-3} - 10^{-2}\ cm^2/(Vs)$ at room temperature. In this study, we present the Hall effect characterization of a-Ge prepared by self-ion implantation of $^{74}Ge^+$ ions. The a-Ge prepared by this method is highly homogenous and has a mass density within 98.5% of the crystalline Ge. The material exhibits an exceptionally high electrical conductivity and carrier mobility ($\sim 100\ cm^2/(Vs)$) for an amorphous semiconductor. The temperature-dependent resistivity of the material is very-well defined with two distinctive regions, extrinsic and intrinsic conductivity, as in crystalline Ge. These results are direct evidence for a largely-preserved band structure and non-localized states of the valence band in a-Ge, as proposed by Tauc *et al.* from optical characterization alone. This finding is not only significant for the understanding of electrical conductivity in covalent disordered semiconductors, but the exceptionally high mobility we have observed in amorphous Ge opens up device applications not previously considered for amorphous semiconductors.

Amorphous semiconductors are an important set of materials with a wide range of applications in thin film solar cells, flat panel displays, thin film electronics and wearable devices [1]. The unique advantages of amorphous semiconductors as compared with their crystalline counterparts include the possibility of producing the materials on many types of substrates (including flexible substrates) with low thermal budget (often at room temperature). Despite having a disordered atomic arrangement, the main features of the electronic band structure are retained in the amorphous phase, including a bandgap quite comparable to the crystalline counterpart. Due to the lack of translational symmetry, theoretical calculation of the electronic band structure of amorphous materials can be very difficult [2]. Nevertheless, the simplified one-dimensional problem has been solved, providing some important features of the disordered materials [3]. Treated as a perturbation from the periodic structure, the interatomic distance of the disordered phase can be formulated as $a(1 + \varepsilon\gamma)$, where $a$ is the lattice constant of the crystalline phase, $\varepsilon$ is a constant for the degree of disorder, and $\gamma$ is a random variable that follows a normalized Gaussian distribution ($\gamma_{av} = 0, \gamma_{av}^2 = 1$). The solution for this one-dimensional case has shown that a bandgap exists but the electronic wave functions become localized at the band extrema and extend into the energy gap, depending on the degree of disorder $\varepsilon$. In this localized band, electrons cannot freely travel in space without exchanging some energy with the surrounding environment, usually with phonons, and jump from one state to another. In other words, electrical conductivity is usually by the hopping mechanism [4]. In the three-dimensional case, however, it might be expected that the localization of the electron wave function could be less likely partly because the electron wave may find ways to travel around the defects. Therefore, non-localized states may still be able to occur at the band extrema [5].

It is a further challenge to reconcile all the available experimental data and to construct a unified band structure for an amorphous material because the specific material properties are particularly dependent on several factors such as the fabrication methods and the sample history. It is reported that the conductivity and the thermoelectric power of amorphous germanium (a-Ge) and silicon (a-Si) prepared by electron-beam evaporation vary quite markedly with the substrate temperature, evaporation rate and the ambient pressure during the evaporation [6]. Ageing effects and the absorption of impurities from the ambient can also change the electrical conductivity, with significant film to film variability [7]. In particular, reports on the type of the conductivity of a-Ge are conflicting as to whether it is p-type or n-type. For example, using Hall effect measurements, Clark *et al.* showed that the conductivity is n-type with a carrier concentration of the order of $10^{18}\ cm^{-3}$ [8], whereas, Grigorovici *et al.* showed that a-Ge has p-type conductivity with a carrier concentration of $\sim 10^{17} - 10^{18}\ cm^{-3}$, regardless of the dopant in the evaporated materials [5,9,10]. Most of the electrical characterization results to date seem to agree that the carrier mobility of deposited a-Ge is less than $\sim 10^{-2}\ cm^2/(Vs)$, which usually leads to a conclusion that the electronic wave function is localized and the conductivity is by the hopping mechanism.

In contrast to electrical characterization, optical characterization techniques provide results that support non-localized states in a-Ge [2]. Studying the absorption spectra of a-Ge from $0.08\ eV$ to $1.6\ eV$, Tauc *et al.* observed optical transitions characteristic of the three branches of the valence band [9]. These transitions are well-described by a set of formulae used to describe the valence band of crystalline Ge (c-Ge). Calculation of the hole effective mass and the spin-orbit splitting energy from the optical results also showed comparable values to crystalline Ge. An important conclusion of the study [9] was that the valence band of the a-Ge is not much different from that of the c-Ge and the wave-function of the valence band is non-localized as in c-Ge [5,9,11]. However, such an extended wave function in a-Ge has not been accepted because the carrier mobility as measured by electrical conductivity is reported to be 4 orders of magnitude lower than what would be expected if the wave function is non-localized. It is noteworthy that all the a-Ge samples in the literature have been prepared by deposition techniques in which the material density varies largely from $72\% - 95\%$ of the c-Ge value [8,12,13]. This reduced density, as well as impurity content, is likely to cause fluctuations in mass distribution in the films, effectively introducing defects, that can affect the properties of the material. Indeed, a more recent report of ion-implanted a-Ge has indicated that it may be possible to prepare a-Ge that has high mobility [14]. However, from this report it was difficult to draw definitive conclusions due to the significant contribution of the c-Ge substrate to the electrical conductivity.

In this letter, we report on the electrical characterization of a-Ge using temperature-dependent Hall effect measurements and demonstrate that our a-Ge is p-type and has a mobility of $\sim 100\ cm^2/(Vs)$. The exceptionally high mobility of this a-Ge is attributed to the non-localized states of the valence band as suggested by Tauc *et al.* [9]. An important difference between this study and almost all of the prior literature is the use of ion implantation to amorphize a crystalline Ge sample, which gives a pure, completely disordered amorphous Ge film with a mass-density very close ( within 98.5%) to the crystalline value [15].

The starting samples before ion-induced amorphization were fully-relaxed crystalline Ge layers grown on Si substrates. It has been known that Si amorphized by ion implantation is a good insulator with a resistivity in the order of $10^5 \, \Omega cm$ [16]. The purpose of using this sample structure is to utilize the insulating property of a-Si for the electrical characterization of the much more conducting a-Ge layer. It will be shown later that the resistance ratio of the a-Si layer over the a-Ge layer is at least $4 - 6$ orders of magnitude. Therefore, the contribution of the a-Si layer to the electrical measurement of the overlaying a-Ge is negligible. The amorphization process was done by implanting $^{74}$Ge$^+$ ions into a (100) Ge-on-Si substrate at an energy of $2.5 \, MeV$ and at a dose of $1 \cdot 10^{15} \, cm^{-2}$. The samples were rotated $7^o$ off the (100) axis to avoid channelling. According to a SRIM simulation [17], these implant parameters are able to induce complete disordering of the original crystal structure with the calculated displacement-per-atom (dpa) > 2 for the ~0.91 $\mu m$ Ge film, and an extension of complete disorder to ~1.1 $\mu m$ in the underlying Si substrate. A previous study has shown that a $dpa$ value of ~0.3 is sufficient for the complete amorphization of a Ge crystal [18]. During the implantation process, the substrates were kept at liquid nitrogen temperature to avoid the formation of porosity in Ge which is a well-known phenomenon for irradiated Ge [19-21]. Low temperature implantation also reduces the probability of self-annihilation of defects and hence ensures full amorphization [22].

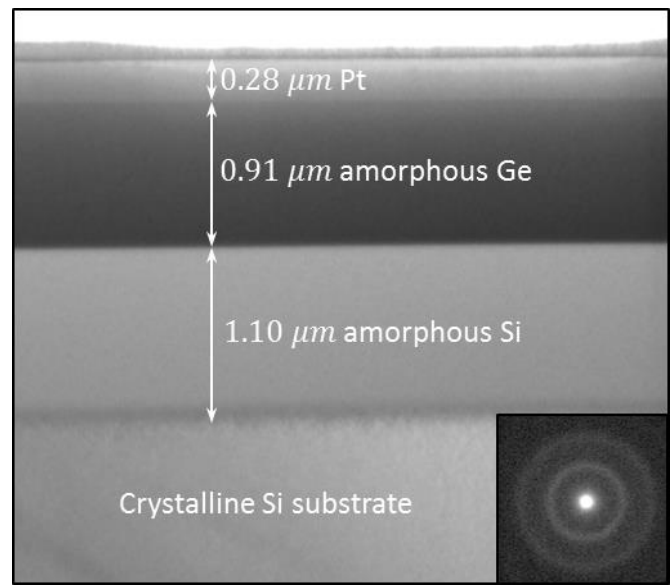

fig. 1: Cross-sectional transmission electron microscopy image of the Ge-on-Si substrate after implanting with $2.5 \, MeV$ $^{74}$Ge$^+$ ions at a dose of $1 \cdot 10^{15} \, cm^{-2}$. Inset is a selected area diffraction pattern of the amorphized Ge layer.

Fig. 1 is a cross-sectional transmission electron microscopy (XTEM) image of the Ge-on-Si sample after ion implantation. The electron-transparent TEM lamella was prepared by ion milling using a focused-ion beam (FIB) system. The top layer of platinum (Pt) was deposited in the FIB chamber before the ion milling to protect the region-of-interest from ion beam damage and sputtering. The structure of the as-implanted samples can be observed to have a $0.91\ \mu m$ a-Ge film, a $1.1\ \mu m$ layer of a-Si and finally the crystalline Si substrate. The a-Ge film is free of voids and is determined to be amorphous as shown by the electron diffraction pattern within the a-Ge layer (inset). Electrical characterization was done using a Lakeshore 7707 Hall effect measurement system. The temperature range of the measurement was from $30K$ to $360K$ using two different measurement stages: a closed cycle refrigerator stage for the temperature range of $30K - 290K$ and a high temperature oven stage for the temperature range of $300K - 360K$. A $1\ cm \times 1\ cm$ Van der Pauw sample structure was used, where electrical contacts to the samples were provided by indium (In) solder bumps. The diameter of the In bumps was kept as small as possible and close to the 4 corners of the samples to minimize measurement errors related to the Van der Pauw structure. I-V measurement of the samples showed good Ohmic behaviour of the contacts. More information on the measurement of the Hall effect is given in the Supplementary section.

In Fig. 2, the resistivity $\rho$ (Fig. 2a) and the Hall coefficient $R_H$ (Fig. 2b) obtained from the measurements are presented. In Fig. 2a, our measured resistivity for the crystalline Ge substrate (blue-triangles) has a similar behaviour with temperature as compared with the measurement from Putley [23]. This curve has two distinctive regions with different slopes. The low-slope region in the low temperature range originates from extrinsic conductivity as a result of dopants (or impurities) in the materials. At a sufficiently high temperature ($\sim 275\ K$), thermal energy can excite electrons from the valence band into the conduction band, exponentially reducing the resistivity. This high-slope region is characterized by intrinsic conductivity involving band-to-band transitions in the materials. It is noticeable that the Hall coefficient of the c-Ge samples (blue-triangles) in Fig. 2b changes from positive in the extrinsic region to negative in the intrinsic region (but is plotted as positive to be compatible with the logarithmic scale). The sign of the Hall coefficient is indicated by empty (negative) and filled (positive) triangles. This can be explained by the fact that the conductivity of the c-Ge substrate is dominated by p-type mono-carriers at low temperature, then exhibits mixed-conductivity with both electrons and holes at higher temperatures. Since the Hall coefficient $R_H$ for the mixed-conductivity depends on both electron and hole mobility, $R_H$ becomes negative at the transition temperature due to the higher mobility of electrons. This carrier type-conversion phenomenon is well known and is usually observed in lightly-doped p-type c-Ge samples and occurs at around room temperature [23].

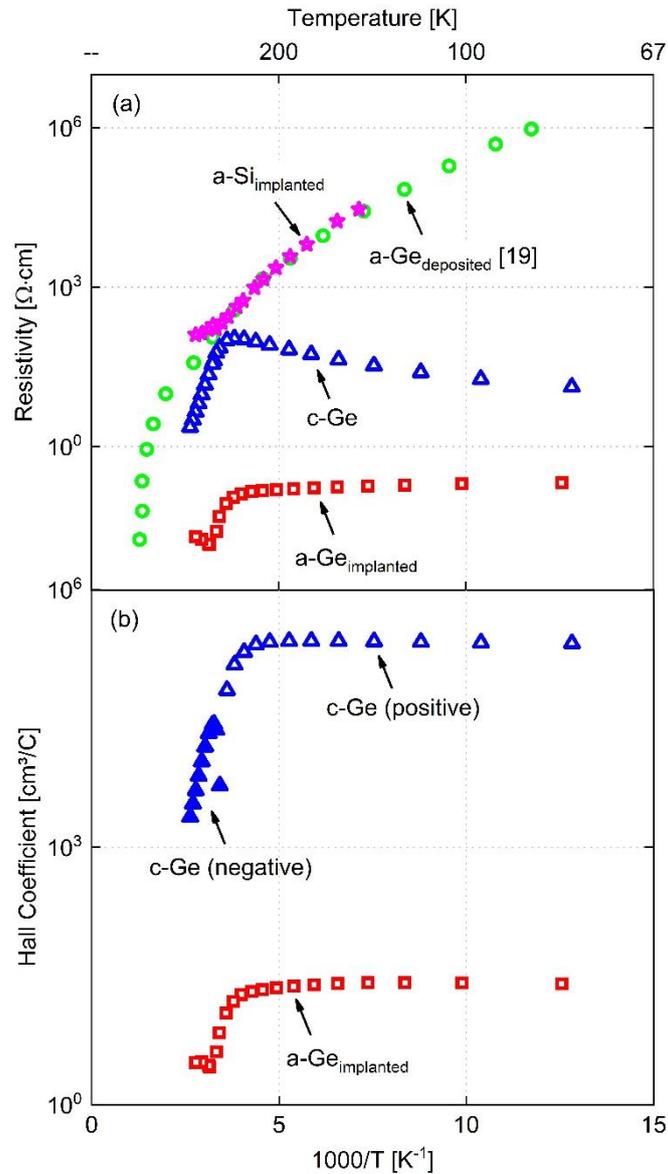

Fig. 2: Resistivity (a) and Hall coefficient (b) as a function of the reciprocal temperature ($1/T$) for c-Ge (blue-triangles), a-Ge by deposition (green-circles) from [24], a-Si by implantation (magenta-stars) and a-Ge by implantation (red-squares). For the Hall coefficient of c-Ge, $R_H$ is positive (empty triangles) when $T < 275\ K$. When $T > 275\ K$, $R_H$ is negative (filled triangle), but is presented to be positive in (b) to be compatible with the logarithmic scale.

For the a-Ge samples, the difference between the measured resistivity of the implanted case (red-squares) and the thermally deposited one (green-circles) is substantial [24]. For the deposited materials, there is a gradual increase in resistivity between $10^3 - 10^6$ Ωcm for $T < 275\ K$. In contrast, the resistivity curve of the implanted a-Ge is very well-defined and has a similar trend to the c-Ge case, with an extrinsic region for $T < 275\ K$ (low T) and an intrinsic region for $T > 275\,K$ (high T). While the band structure of the deposited a-Ge has been previously referred to as "fuzzy" due to localized band tails at the top of valence band and the bottom of the conduction band [24], the band structure of our implanted a-Ge, based on the resistivity data, is clearly more well-defined, closely resembling that of its crystalline counterpart. At low temperature, the resistivity of the implanted a-Ge is almost constant at around $10^{-1}\,\Omega cm$, which is $4 - 7$ orders of magnitude lower than the typical values for deposited a-Ge. As indicated by the Hall coefficient in Fig. 2b, the conductivity of the implanted a-Ge is p-type since $R_H$ is positive.

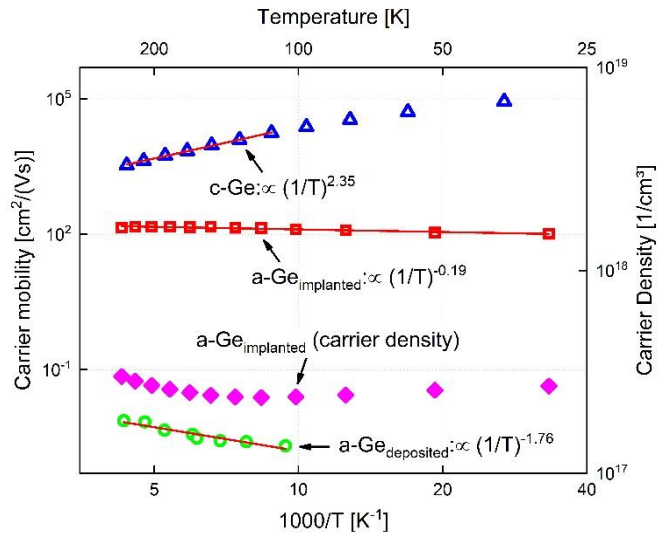

Fig. 3: Carrier mobility as a function of reciprocal temperature $(1/T)$ for c-Ge (blue-triangles), a-Ge by deposition (green-circles) [25] and a-Ge by implantation (red-squares). Magenta-diamond is the carrier density of the a-Ge by implantation.

From the measured resistivity and the Hall coefficient data, it is quite straightforward to calculate the carrier mobility and the carrier density for the extrinsic conductivity region ($T < 275K$), noting that we are not examining the intrinsic region ($T > 275K$) in this report. In the extrinsic region, the conductivity is dominated by a single carrier for both crystalline and implanted a-Ge because the resistivity of these two materials is almost constant with temperature. For a single carrier conductivity, the carrier density, $p$, can be found from:

$$p = \alpha \left(\frac{1}{R_H \cdot e}\right),$$

where $e$ is the elementary charge ($1.6 \times 10^{-19}\,C$), $\alpha$ is a constant but in general can be ignored since it is always close to unity, and, in the worst case, it will introduce only a small error in the carrier concentration [23]. Similarly, the carrier mobility can be found from:

$$\mu = \frac{R_H}{\rho}$$

Extracted data for the carrier density $n$ and the carrier mobility $\mu$ of c-Ge and a-Ge are shown in Fig. 3. The carrier concentration of the implanted a-Ge (magenta-diamonds) is of the order of $10^{17}\ cm^{-3}$ and is p-type. The carrier concentration remains largely constant within the temperature range, which means that most acceptor levels are very close to the valence band edge. Tauc *et. al* estimated the activation energy of such acceptors as $E_A \geq 0.01\ eV$ [9].

Treating c-Ge first, the mobility of c-Ge (blue-triangles) in Fig. 3 is in excellent agreement with studies by Morin *et. al* [26,27]. In general, the carrier mobility depends on impurity and lattice scattering. However, impurity scattering in high quality materials decreases as the temperature increases and is negligible for $T > 100\ K$. Due to acoustic lattice scattering, the carrier mobility should exhibit a temperature dependence given by: $\mu_{ac} = AT^{-1.5}$. From Morin's study, experimental data of c-Ge in the range of $100K - 300K$ showed a temperature dependence of $T^{-1.66}$ for electrons and $T^{-2.33}$ for holes, deviating from the $T^{-1.5}$ law [26]. Our experimental data in Fig. 3 exhibits a dependence of $T^{-2.35}$ for holes, which closely matches Morin's value. By including optical lattice-scattering, Morin *et. al* were able to correct the deviation for electrons, but not for holes. These authors attributed this to the fact that the valence band minimum is not at the center of the Brillouin zone [26,27].

For the a-Ge case, the carrier mobility of the evaporated amorphous material is in the order of $10^{-3}\ cm^2/(Vs)$ (green-circles), typical of values found in the literature for a-Ge [25]. According to a theoretical calculation, one would expect the Hall mobility in the variable-range-hopping regime to be around $10^{-4}\ cm^2/Vs$ [28]. Remarkably, the mobility of a-Ge produced by implantation is found to be exceptionally high, of the order of $10^2\ cm^2/(Vs)$ (red-squares). At this moment, it is worth re-visiting the works of Tauc *et. al* [9] who reported a number of findings on the band structure of deposited a-Ge. The absorption spectra at photon energies of $0.08 - 0.6\ eV$ showed several direct transitions between the three branches of the valence band, similar to what is observed in p-type c-Ge. Indeed, the absorption spectra of the a-Ge can be well described by the same set of formulae used for c-Ge, whereby the effective masses of heavy holes $m_1$, light holes $m_2$ and the energy gap of the spin-orbit splitting $\Delta E$ are comparable to those for c-Ge. Hence, it was suggested that the valence band is essentially preserved in a-Ge and the wave function can be described by Bloch functions giving a non-localized wave. Based on previous electrical data for deposited a-Ge, this proposal by Tauc *et. al* was not accepted by others [8,24,29,30]. In contrast, our measurements for implanted a-Ge show a very well-defined behaviour of the resistivity with temperature (Fig. 2) and the carrier mobility is extremely high for an amorphous semiconductor, of the order of $10^2\ cm^2/(Vs)$ (Fig. 3). Hence, these two findings provide direct evidence in favour of Tauc's suggestion for the non-localized states of the valence band. It is worth noting that our measurement of implanted a-Si showed a resistivity of $10^2 - 10^4\ \Omega cm$ (Fig. 2a/magenta-stars), which is $4-6$ orders of magnitude higher than that of the implanted a-Ge, indicating a fundamentally different conductivity mechanism between a-Si and a-Ge. While a-Si can be characterized by hopping conductivity, carrier transport in pure ion implanted a-Ge is largely non-localized.

To account for the large difference in electrical results measured in this study and the previous data for deposited a-Ge, we propose that it is not due to the intrinsic nature of a-Ge, but mostly dictated by density fluctuations within deposited materials, which effectively introduce defects. This proposal was first adopted by Stuke for amorphous selenium [31] and later adopted by Tauc for a-Ge [11]. The density fluctuation in the materials may cause different types and concentrations of defects such as dangling bonds and different bonding configurations to the four-fold coordinated case expected for a-Ge, internal strain or introduce defect related carriers, all of which can contribute to potential barriers to carrier transport. Tauc speculated that a potential barrier of $\sim 0.1\ eV$ would be sufficient for the 4 orders of magnitude difference between the conductivity of the deposited a-Ge, compared with the extended electron wave functions that he proposed for a-Ge. Overall, our findings for implanted a-Ge provide a pathway for producing a high mobility a-Ge material that behaves largely like c-Ge with a similar band structure and non-localized states. Improved deposition methods for a-Ge with low impurity content and a density close to that of ion implanted a-Ge should provide a homogeneous film that can lead to similar results and hence to significant advances in thin film electronics, solar cells or flexible devices.

In summary, this study has investigated the electrical properties of a-Ge prepared by implantation of $^{74}Ge^+$ ions. The material has a significantly higher mass density and homogeneity as compared to a-Ge prepared by deposition methods. Unlike previous studies of deposited a-Ge, the temperature-dependence of the resistivity is very well defined, indicating a well-defined band structure in implanted a-Ge. In addition, the hole mobility is exceptionally high ($\sim 10^2\ cm^2/(Vs)$) for an amorphous material. These results provide direct evidence for a largely preserved valence band and a non-localized, extended electron wave function in high quality a-Ge, as previously proposed by Tauc *et. al* based on optical measurements alone.